\documentclass[iop,apj]{emulateapj}
\usepackage{apjfonts}
\usepackage{epsf}
\newcommand{\Msun}{\mathrm{M}_{\odot}}
\newcommand{\Msunyr}{\Msun\ \mathrm{yr}^{-1}}

\def\dim#1{\mbox{\,#1}}

\begin{document}

\slugcomment{Submitted to ApJ}
\shortauthors{Muratov et al.}
\shorttitle{The effects of Population III stars on their host galaxies}

\title{Revisiting The First Galaxies: The effects of Population III stars on their host galaxies} 
 
\author{Alexander L. Muratov \altaffilmark{1}, Oleg Y. Gnedin \altaffilmark{1}, Nickolay Y. Gnedin\altaffilmark{2,3,4}, Marcel Zemp \altaffilmark{1,5}}

\altaffiltext{1}{Department of Astronomy, University of Michigan, 
   Ann Arbor, MI 48109; \mbox{\tt muratov@umich.edu}}
\altaffiltext{2}{Particle Astrophysics Center, Fermi National Accelerator Laboratory, Batavia, IL 60510, USA}
\altaffiltext{3}{Kavli Institute for Cosmological Physics and Enrico Fermi Institute, The University of Chicago, Chicago, IL 60637 USA}
\altaffiltext{4}{Department of Astronomy \& Astrophysics, University of Chicago, Chicago, IL 60637 USA}
\altaffiltext{5}{Kavli Institute for Astronomy and Astrophysics, Peking University, Yi He Yuan Lu 5, Hai Dian Qu, Beijing 100871, China }

\date{\today}
\begin{abstract}
We revisit the formation and evolution of the first galaxies using new hydrodynamic cosmological simulations with the ART code. Our simulations feature a recently developed model for $H_{2}$ formation and dissociation, and a star formation recipe that is based on molecular rather than atomic gas. Here, we develop and implement a recipe for the formation of metal-free Population III stars in galaxy-scale simulations that resolve primordial clouds with sufficiently high density. We base our recipe on the results of prior zoom-in simulations that resolved the protostellar collapse in pre-galactic objects. We find the epoch during which Pop III stars dominated the energy and metal budget of the first galaxies to be short-lived. Galaxies which host Pop III stars do not retain dynamical signatures of their thermal and radiative feedback for more than $10^8$ yr after the lives of the stars end in pair-instability supernovae, even when we consider the maximum reasonable efficiency of the feedback. Though metals ejected by the supernovae can travel well beyond the virial radius of the host galaxy, they typically begin to fall back quickly, and do not enrich a large fraction of the intergalactic medium. Galaxies with total mass in excess of $3 \times 10^6 \, \Msun$ re-accrete most of their baryons and transition to metal-enriched Pop II star formation. 
\end{abstract}

\section{Introduction} 

A natural consequence of Big Bang Nucleonsynthesis is that the first stars in the universe formed from gas that was completely devoid of elements heavier than lithium. Since such conditions do not exist in any known star-forming regions today, star formation in the primordial regime has thus far only been explored by theory and simulations. From first principles, it can be deduced that gas which is free of metals would not be able to cool efficiently, and would therefore inherently have a higher Jeans mass than stars forming in metal-enriched gas. \citet{bromm_etal99} showed that metal-free gas settles into disks, then fragments into clumps with $M_J \approx 10^3 \, \Msun$, which undergo runaway collapse to densities of $n_H > 10^8 \dim{cm}^{-3}$. In a follow-up study, it was shown that this process was robust to initial conditions in Smoothed Particle Hydrodynamics (SPH) simulations \citep{bromm_etal02}. Using independent Adaptive Mesh Refinement (AMR) grid techniques, \citet{abel_etal00} confirmed that such dense clumps and cold pockets can indeed form in primordial gas. \citet{abel_etal02} presented even more realistic simulations, and concluded that single massive stars would form via $H_2$ cooling at the center of these clumps, and the radiative feedback would halt accretion onto the star and prevent further star formation in the parent halo. Subsequent work with higher resolution reaffirmed these conclusions \citep{yoshida_etal06, oshea_norman07}. However, recent studies with both SPH \citep{stacy_etal10, stacy_etal12, clark_etal11} and grid techniques \citep{turk_etal09, greif_etal11} have suggested that angular momentum imparted on the gas during collapse could still lead to fragmentation, causing the cores and the resulting stars to be substantially smaller. Ultimately, the next generation of super-zoomed simulations will need to follow the proto-stellar systems for longer periods of time, with more detailed treatments of radiative transfer and magnetohydrodynamics, to conclude decisively on the true nature of Pop III stars \citep{greif_etal12}.

Though we do not yet have any direct observational evidence for Pop III stars, we know that they must have existed in some form, as gas in the universe inevitably transitioned from having primordial composition to being enriched with enough metals to allow present-day star formation to commence. The nature of this transition is of key importance for understanding the dawn of galaxy formation: if Pop III stars did indeed form with a top-heavy initial mass function (IMF), a significant fraction of them may have been prone to end their lives as pair-instability supernovae (PISNe). Such supernovae generate up to ten times as much thermal energy as Type II SNe \citep{heger_woosley02}, quickly heating the gas in their host halos. In addition, metal-free stars are able to produce enormous quantities of ionizing photons: a metal-free star will always have a higher surface temperature than an enriched counterpart of equal mass \citep{schaerer02, tumlinson_shull00}. If the IMF is indeed top-heavy, the effect is even more drastic as many stars would have their emission spectra peak in the UV regime. Both the supernovae and the ionizing photons serve to heat and disperse neutral gas in the vicinity the star, creating a large HII region \citep{whalen_etal04, alvarez_etal06, johnson_etal07,abel_etal07}, and delaying the onset of steady, continuous star formation. On the other hand, PISNe release a large amount of metals, rapidly enriching previously primordial gas \citep{wise_abel08, greif_etal10, maio_etal11, wise_etal12}, and potentially leading to quick termination of the Pop III epoch \citep{yoshida_etal04}.
The balance of these effects works to determine star formation rates in galaxies and their subsequent evolution. 

While Pop III stars are no longer thought to be a major driver of the global reionization of the intergalactic medium \citep{ciardi_etal00, ricotti_etal02b, ricotti_ostriker04, mesinger_etal09}, constraints on the cosmological electron scattering optical depth from WMAP suggest that halos less massive than $ 10^8 \, \Msun$ may have contributed to the photon budget at the beginning of reionization. In order for this scenario to work, low mass halos must permit the escape of ionizing photons effectively \citep{alvarez_etal12, ahn_etal12}. Pop III stellar feedback has been explored as a mechanism to create windows of time during which such high escape fractions are made possible (\citealt{ wise_cen09}, but see \citealt{gnedin_etal08,ricotti_etal08}). 

We revisit these important conclusions with novel high-resolution cosmological simulations that feature separate star formation criteria and feedback prescriptions for Pop III and Pop II stars. While we cannot resolve all stages of their formation, we do resolve the clumps of gas on a \textasciitilde 1 pc scale which inevitably collapse into Pop III stars. In addition to the commonly accepted parameters for Pop III formation and feedback, we also explore models in which the radiative and SNe feedback are taken at extreme values, as well as models in which the IMF of Pop III stars is taken to be identical to Pop II counterparts. 

Our analysis in this paper focuses on the dynamical effects that Pop III stars can impart onto their host galaxies. We quantify the ability of Pop III stars to suppress star formation, expel gas, and enrich the medium both within and outside of the galaxies in which they form. Using a wide suite of simulations, we show that mass resolution and mesh refinement criteria affect the derived importance of Pop III stars. In a forthcoming second paper (\citealt{muratov_etal13}, Paper II), we will explore the nature of the transition between Pop III and Pop II star formation, and assess the relative importance of feedback effects from the two stellar populations.

\section{Simulations}
\label{sec:sims}
We perform cosmological simulations with the Eulerian, gasdynamics+N-body adaptive refinement tree (ART) code \citep{kravtsov_etal97, kravtsov99, kravtsov03, rudd_etal08}. The latest version of the code incorporates a new phenomenological prescription for molecular hydrogen formation on dust grains and self-shielding, as well as shielding by dust introduced in \citet{gnedin_etal09} and developed further in \citet{gnedin_kravtsov11}. Having such a detailed account of molecular gas, as well as excellent spatial resolution at high redshift makes it practical to consider a star formation recipe that is also based on molecular gas. This formulation has previously been shown to much better reproduce the Kennicutt-Schmidt relation for high-redshift, low-mass, and low metallicity galaxies, and it now enables more realistic simulations of the early universe (\citealt{tassis_etal08, gnedin_kravtsov10, gnedin_kravtsov11}). Radiative transfer, including Hydrogen and Helium ionization, as well as Lyman-Werner $H_2$ dissociating feedback, is computed using the OTVET approximation \citep{gnedin_abel01}, employing the same Eddington tensor considered in that work. Stellar particles are treated as point sources of radiation. Diffuse radiation from the CMB, Compton heating, recombination, and bremsstrahlung is also computed.

For our baseline runs, we use a 1 $h^{-1}$ Mpc comoving box and the WMAP-7 cosmology ($\Omega_m = 0.28$, $\Omega_\Lambda = 0.72$, $h=0.7$, $\sigma_8 = 0.817$, $\Omega_{b} = 0.046$, $\Omega_{DM}=0.234$). Additional runs were performed with 0.5 $h^{-1}$ and 0.25 $h^{-1}$ Mpc comoving boxes to explore the effects of mass resolution (hereby referred to as H Mpc and Q Mpc runs, respectively). The details for each run performed in our simulation suite are presented in Section \ref{sec:tests}. Numbers quoted in this paragraph, as well as Sections \ref{sec:popII} and \ref{sec:pop3sf} refer to our baseline 1 $h^{-1}$ Mpc runs. These runs start with a $256^3$ root grid, which sets the DM particle mass to $m_{\, DM} = 5.53 \times 10^3 \, \Msun$ and the base comoving resolution of 5.56 kpc. We employ Lagrangian refinement criteria, refining cells when the DM mass approximately doubles compared to the initial mass in the cell, specifically, when it exceeds $2 \times m_{\, DM} \times \frac{\Omega_m}{\Omega_{DM}} \times 0.8$, or the gas density surpasses an approximately equivalent value modulated by the cosmic baryon fraction $0.3 \times m_{\, DM} \times \frac{\Omega_m}{\Omega_{DM}} \times 0.8$. In both refinement conditions, the extra factor of $0.8$ is the split tolerance. We use a maximum of 8 additional levels of refinement, giving us a final resolution of $10^6 h^{-1} / 256 / 2^8 \approx22$ comoving pc. Since we are studying high-redshift galaxies, it is important to note that this translates to about 2 physical pc at the endpoint of our simulations, $z=9$, and 1 physical pc at $z=20$ when the first stars begin to form. This spatial resolution is sufficient to capture the detailed multi-phased structure of the interstellar medium (ISM) (e.g. \citealt{ceverino_klypin09}).

In order to simulate several representative regions of the universe, we employ the "DC mode" formulation presented in \citet{sirko05} and \citet{gnedin_etal11}. Running simulations with different DC modes allows us to sample cosmologically over- and under-dense regions without actually changing the total mass within each box. A single parameter $\Delta_{DC}$, which is constant at all times for a given simulation box, represents the fundamental scale of density fluctuations present in the box. At sufficiently early times, when perturbations on the fundamental scale of the box are in the regime of linear growth, $\Delta_{DC}$ is related to the overdensity, $\delta_{DC}(a) \equiv D_+(a)\,\Delta_{DC}$, where $D_+$ is the linear growth factor. The expansion rate of the individual box relates to the expansion rate of the universe by the following relation, which is Equation 3 of \citet{gnedin_etal11}:
\begin{equation}
a_{box} = \frac{a_{uni}}{\left[ 1 + \Delta_{DC}D_+(a_{uni})\right]^{1/3}},
\end{equation}
where $a_{box}$ is the local scale factor of the simulation box, while $a_{uni}$ is the global expansion factor of the universe. For this study, we have used three different setups with $\Delta_{DC} = -2.57, -3.35$, and $4.04$, labeled 'Box UnderDense$-$', 'Box UnderDense+' and 'Box OverDense', respectively. At the endpoint of our simulations, $z=9$, these values translate to overdensities of $\delta_{DC} = -0.257, -0.335$, and $0.404$, respectively. Boxes UnderDense$-$ and UnderDense+ have negative DC modes, implying they are underdense regions of the universe. However, while Box UnderDense$-$ is representative of a void, and hosts only low-mass galaxies which collapse relatively late, Box UnderDense+ hosts the first star-forming galaxy among all three simulation boxes. This galaxy is also more massive than any of those in Box OverDense until $z\approx13$.

Box OverDense hosts several massive star-forming galaxies which statistically dominate the sample of simulated galaxies. The H Mpc and Q Mpc boxes used $\Delta_{DC} = $ 5.04 and 6.11, respectively. These runs are primarily designed to explore the earliest possible epoch of Pop III star formation, tracing only the most overdense regions with even higher mass resolution. None of our simulations continue past $z=9$, as the boxes are too small to capture the nonlinear growth of large-scale modes at later epochs.

We construct catalogs of halo properties from simulation outputs using the profiling routine described in \citet{zemp_etal12}. We take the virial radius, $R_{vir}$, as the distance from the center of a halo which encloses a region that has an overdensity of 180 with respect to the critical density of the universe. 

\subsection{Population II star formation}
\label{sec:popII}
Following \citet{gnedin_etal09} and \citet{gnedin_kravtsov11}, we set the threshold for Pop II star formation in a gas cell when the fraction of molecular hydrogen exceeds the threshold $f_{H_2} \equiv 2n_{H_2} / n_H = 0.1$. Tests described in the above studies showed that the exact value of this threshold was not important for overall star formation rates, but mainly regulated the number and mass of stellar particles produced. The simulations performed in that study employed a non-zero floor (minimum value) of the dust-to-gas ratio in cells, which was meant to account for unresolved pre-enrichment. Since our current simulations spatially resolve regions where the first stars are expected to form, it is unnecessary and inappropriate to use such a floor. This means that in our runs, prior to the metal feedback from the first generation of stars, only primordial chemistry is used in $H_2$ formation reactions. We find that such primordial reactions with our resolution do not yield molecular fractions $f_{H_2}>0.01$ on relevant timescales. Therefore, in order to form the first (Pop III) generation of stars, a separate prescription is required and is outlined in the next section. 

In cells where the molecular fraction exceeds the $f_{H_2}$ threshold, Pop II stellar particles are formed with a statistical star formation delay, $dt_{SF} = 10^7$ yr, implemented by drawing a random number, $P$, between 0 and 1, and forming stars only if $P>\exp{\left(-\frac{dt}{dt_{SF}}\right)}$, where $dt$ is the length of the timestep at the cell's refinement level. Each particle represents a stellar population with a \citet{miller_scalo79} IMF from 0.1$\, \Msun$ to 100 $\, \Msun$. The mass of a stellar particle is determined by the following the relation:
\begin{equation}
\dot{\rho_*} = \frac{\epsilon_{ff}}{\tau_{SF}} \rho_{H_2}.
\label{eq:SFC}
\end{equation}
The star formation efficiency per free-fall time is set to $\epsilon_{ff}=0.01$, based on recent results of \citet{krumholz_etal12}. We use a constant star formation timescale $\tau_{SF} = 8.4 \times 10^6$ yr corresponding to the free-fall time at hydrogen number density $n_H = 50 \dim{cm}^{-3}$. This approach differs from \citet{gnedin_kravtsov11}, where the timescale was computed by using the physical density of molecular clouds. However, we find that in our simulations the density-dependent timescale instills a strong resolution dependence on the star formation rate. 
Pop II stellar particles are treated as statistical ensembles of stars for which the appropriate metal yield and fraction of stars to go supernovae is computed by integration of the IMF. The number of SN II explosions is $75$ per $10^4 \, \Msun$ formed, while the amount of SN II metals generated by a stellar particle is $1.1 \%$ of its initial mass. Each supernova releases $2 \times 10^{51}$ erg thermal energy which is deposited over the course of $10^7$ yr. Following the notation of \citet{hummels_bryan12}, this implies fraction of the rest mass energy of stars which is available for thermal SNe feedback is $E_{SN} / Mc^2 = 8.4\times10^{-6}$. This value is relatively high, but consistent with values chosen by other researchers \citep{hummels_bryan12}.

\subsection{Population III star formation}
\label{sec:pop3sf}
\subsubsection{Formation criteria}
We model the formation of Pop III stars based on criteria derived from simulations of \citet{abel_etal02}. These authors showed that once the core density of a proto-cloud reached 1000 $\dim{cm}^{-3}$, further collapse to a massive stellar object was imminent. Analyzing their results, we found that for gas at any given density $n_H$ past this threshold, the time of collapse to a stellar core is approximately six times the free-fall time for that density, $6t_{ff}(n_H)$. This collapse time is 9 Myr for $n_H = 1000 \dim{cm}^{-3}$ and scales as $n_H^{-1/2}$. For our fiducial runs, we use $n_{H, min} = 10000 \dim{cm}^{-3}$ as the threshold and $dt_{SF}= 2.8$ Myr as the statistical star formation delay, simulating the collapse time. This value is lower than the one used for Pop II stars. This density threshold value was chosen to ensure Pop III stars would form primarily when cells have been maximally refined, but is low enough such that the collapsing gas clouds are still fully resolved in our simulations. Further discussion is presented in Section \ref{sec:nHtest}.

We also set a threshold for the minimum fraction of molecular hydrogen at $10^{-3}$ to reflect that primordial gas clouds must cool primarily via ro-vibrational transitions of $H_2$ to form the first stars \citep{couchman_rees86, tegmark_etal97}. The precise value of this threshold is rather arbitrary, as we do not attempt to model the actual chemistry of stellar core formation. We have chosen this value because it is lower than, but close to the typical value for the molecular hydrogen fraction in cold, dense primordial gas around $z=20$, which we have found empirically to be $2 \times 10^{-3}$ (see Section \ref{sec:molfrac} and Figure \ref{fig:fH2nH}). A minimum threshold for molecular fraction ensures that the $H_2$-dissociating Lyman Werner radiation from recently-formed Pop III stars will realistically suppress further Pop III star formation in the region.

Pop III stars form in gas that has metallicity $\log_{10} Z/Z_\odot < -3.5$. This threshold is chosen to match the critical metallicity discovered by \citet{bromm_etal01}, and has held up in later studies \citep{smith_etal09}. Though the exact value of this critical metallicity is still uncertain and can be affected by the presence of dust \citep{omukai_etal05}, we find that it is not very important as the majority of Pop III stars form in truly primordial, or nearly primordial gas. Compiling all of our simulations, we found that only \textasciitilde10\% of Pop III stars form with $\log_{10} Z/Z_\odot > -5$. 

We summarize the formation criteria for Pop III stars with the following set of equations, \vspace{-0.1cm}
\begin{eqnarray}
      && n_H > n_{H, min} = 10^4 \dim{cm}^{-3}
      \nonumber\\ && f_{H_2} > f_{H_2, min} = 10^{-3}
      \nonumber\\ && \log_{10} Z/Z_\odot < -3.5. 
  \label{eq:criteria}
\end{eqnarray}
Values given for each variable represent the fiducial choices.

\subsubsection{IMF and supernova feedback}

The IMF of Pop III stars is currently a hotly debated and active area of research. It is still unclear whether the high Jeans mass of primordial gas results in a top-heavy IMF as predicted by early studies \citep{abel_etal02, bromm_etal99, yoshida_etal03}, or if the angular momentum and radiative effects during infall can fragment the cloud and generate relatively low-mass cores \citep{greif_etal11, stacy_etal12, hosokawa_etal11, clark_etal11}. It is even likely that the Pop III IMF can be considerably variable depending on environment \citep{oshea_norman07} and ionization state of the collapsing gas \citep{yoshida_etal07}. We choose not to explore various analytic forms for the IMF, as constraining it is beyond the scope of this paper. Instead, we consider that the main way by which the Pop III IMF can influence galaxy formation, in contrast to the known Pop II IMF, is by enhancing the output of ionizing radiation and the number and intensity of supernovae. In particular, PISNe, which are hypothetically plausible from stars in the mass range $140 - 260 \, \Msun$ (or for lower masses if rotation is considered, see e.g. \citealt{stacy_etal13}), would potentially be dramatic singular events in the evolution of any galaxy \citep{bromm_etal03, whalen_etal08}. To account for the occurrence of PISNe we use two different particle masses for Pop III stars. Every newly formed Pop III stellar particle is randomly assigned to be either a $170 \, \Msun$ star, which is to explode in a PISN, or $100 \, \Msun$ star, which only generates a mild explosion before collapsing into a black hole \citep{heger_woosley02, heger_woosley10}. The proportion of these two types of particle mass and fate is governed by a single parameter, $P_{PISN}$, which is the fraction of PISNe progenitors ($170 \, \Msun$ stars) that form when the Pop III star formation criteria are met. In our fiducial runs, we set $P_{PISN} = 0.5$. This value was chosen to test the maximum possible impact of PISNe on galaxy evolution, and probably represents the most top-heavy the primordial IMF can possibly be. Since the atmospheres of Pop III stars are free of metals, they are unable to drive stellar winds and therefore do not enrich the ISM in any way other than through supernovae. Pop III stars which have masses too low to produce SNe are ignored in our model.

For our fiducial value of $n_{H, min}$ (as well as all other parameters considered in Section \ref{sec:nHtest}), we found that the gas mass in a maximally refined cell at $z\approx20$ is sometimes not sufficient to form a $170 \, \Msun$ star. Therefore, we prevent further refinement in metal-free cells that have $n_H > 0.5 n_{H, min}$ and whose splitting would leave insufficient mass to form the star. Through tests, we have checked that this refinement restriction never artificially slows down Pop III star formation. It becomes especially relevant in the super-Lagrangian runs discussed in Section \ref{sec:SLR} and in the H and Q Mpc boxes which inherently have very high resolution.

The PISN from a $170 \, \Msun$ star releases $27 \times 10^{51}$ erg of thermal energy, as well as $80 \, \Msun$ of metals into the ISM \citep{heger_woosley02}. As suggested by \citet{wise_abel08}, we use a delay of 2.3 Myr from the formation of a $170 \, \Msun$ particle to its PISN event, representing the main sequence lifetime of this type of star \citep{schaerer02}. After the supernova goes off, the cell which hosts it often winds up with super-solar metallicity. The cooling functions employed in our code are not accurate for these high-temperature high-metallicity conditions associated with the early phases of supernova remnants. We found that while the blastwave expanded regardless of whether or not cooling was turned on, the inner regions of the supernova remnant overcooled. We therefore turned off all metal cooling for gas at temperatures higher than $10^4$K. According to the models of \citet{heger_woosley02}, a $170 \, \Msun$ star is completely disrupted by its PISN event, and all gas mass from the stellar interior would be ejected into the ISM leaving no remnant. The ejecta then consists of $80 \, \Msun$ of metals from the core, as well as the primordial envelope which is $22 \, \Msun$ of He and $68 \, \Msun$ of H.

A $100 \, \Msun$ star does not explode as a PISN, but its actual fate is still uncertain and depends on the details of the stellar rotation and magnetic structure \citep{heger_woosley10}. Before undergoing core collapse, such a star would experience thermonuclear pair-instability pulses that would eject the outer layers of H and He, with possible traces of the elements C, N, and O. The energy released in such pulses is of the order or smaller than the energy of a normal SN type II. If the remaining core has enough rotation to trigger a gamma-ray burst in the collapsar model \citep{woosley_heger12}, it may then lead to a powerful explosion with over $10^{52} \dim{erg}$ of energy and the ejection of significant mass of metals. Without rotation, the core may drive a weak collapsar explosion or no explosion at all, when all the remaining mass recollapses. Given these uncertainties, and to contrast with the case of a full PISN explosion for the $170 \, \Msun$ stars, for the $100 \, \Msun$ stars we assume that no substantial metals are deposited into the ISM and that the released energy corresponds to a standard SN type II. About $50 \, \Msun$ of gas is released into the ISM, while also leaving behind a remnant black hole of \textasciitilde $50 \, \Msun$. 

To prevent artificial radiative losses, PISN energy and mass ejecta are distributed within a sphere of constant density, with a radius 1.5 cell lengths, centered at the middle of the PISN host cell. Each of the 27 cells within such a sphere, consisting of the star's host cell and its immediate neighbors, receive a dose of energy and metal-rich gas proportional to the actual volume of the cell contained within the sphere. This prescription is physically motivated as we found that our typical timestep (about 450 yr) is too coarse compared to the typical timescale of the early free expansion phase of the SN remnant. For example, it takes the ejecta \textasciitilde 500 yr to traverse half of the typical Pop III star host cell length (4.5 pc) at $z$\textasciitilde$20$ if it travels at the free-expansion velocity. This velocity is computed here by assuming that all of the PISN energy of $27 \times 10^{51}$ erg goes into kinetic energy of the ejecta. In practice, we found that this model did not significantly affect the geometry of the blastwave relative to simulations where we injected the metals and energy into a single cell.

\subsubsection{Radiative feedback}
In addition to the supernova feedback, all Pop III stars have enhanced radiative feedback relative to Pop II counterparts, due to the lack of metals in their atmospheres \citep{schaerer02}. We use the same spectral shape for the ionizing feedback of all stellar particles (the Pop II SED from Figure 4 of \citealt{ricotti_etal02a}), which has a characteristic energy of 21.5 eV for ionizing photons, however we enhance the radiative output of Pop III stars by a factor of 10 relative to Pop II, following \citet{wise_cen09}. After a Pop III stellar particle undergoes supernova, radiative feedback from the star is completely shut off. On the other hand, Pop II stellar particles shine according to a light curve fit from Starburst 99 model results \citep{leitherer_etal99}. This light curve consists of a flat component for the first $3\times10^6$ yr, followed by a steep power-law falloff. Radiative feedback from Pop II stellar particles becomes insignificant after $3\times10^7$ yr. Since Pop II stars shine longer than both types of Pop III stars, the factor of $10$ radiative enhancement does not translate into a proportionally higher number of ionizing photons per lifetime. Pop II stars emit 6,600 ionizing photons per stellar baryon per lifetime. In our fiducial runs, Pop III stars emit 38,800 and 34,500 photons per baryon per lifetime for the $100 \, \Msun$ and $170 \, \Msun$ stars, respectively.

\subsection{Convergence Study \& Setting Fiducial Parameters}
\label{sec:tests}

In this section, we describe the test runs that justify the numerical setup and the choice of parameters for our main runs. Since the Pop III star formation recipe described above is one of the critical components of our study, we focus on testing the key elements of this model. In Table \ref{tab:sims} we list the details of the simulations performed in our suite. Box OverDense has many more potential sites for Pop III star formation than the other 1 $h^{-1}$ Mpc boxes, and therefore serves as the best testing ground. It is important to keep in mind that while every parameter we test has an effect on Pop III star formation, the most drastic differences between the simulations are caused by the choice of initial conditions. The role of cosmic variance will be explored more comprehensively in Paper II.

\begin{table*}
\begin{center}
\caption{\sc Simulation runs}
\label{tab:sims}
\begin{tabular}{lrrrrrl}
\tableline\tableline\\
\multicolumn{1}{l}{Run} &
\multicolumn{1}{l}{Base grid} &
\multicolumn{1}{l}{$\ell_{max}$} &
\multicolumn{1}{l}{dx (pc)}  &
\multicolumn{1}{l}{$m_{\, DM} (\Msun)$}  &
\multicolumn{1}{l}{$n_{H, min}($\hspace{-0.05cm}$\dim{cm}^{-3})$}  &
\multicolumn{1}{l}{Description}  
\\[2mm] \tableline\\
\vspace{0.4cm}
\it{Convergence study} \\
\vspace{-0.8cm}
\\[2mm] \tableline\\
UnderDense$-$\_noSF\_7 & $256^3$ & 7 & 44 & 5500 &  & No star formation \\
UnderDense$-$\_noSF\_8 & $256^3$ & 8 & 22 & 5500 &  & No star formation \\
UnderDense$-$\_noSF\_9 & $256^3$ & 9 & 11 & 5500 &  & No star formation \\
UnderDense+\_noSF\_7 & $256^3$ & 7 & 44 & 5500 &  &  No star formation \\
UnderDense+\_noSF\_8 & $256^3$ & 8 & 22 & 5500 &  &  No star formation \\
UnderDense+\_noSF\_9 & $256^3$ & 9 & 11 & 5500 &  &  No star formation \\
OverDense\_noSF\_7 & $256^3$ & 7 & 44 & 5500 &  & No star formation \\
OverDense\_noSF\_8 & $256^3$ & 8 & 22 & 5500 &  & No star formation \\
OverDense\_noSF\_9 & $256^3$ & 9 & 11 & 5500 &  & No star formation \\
\vspace{0.4cm}
OverDense\_noSF\_HMpc & $256^3$ & 8 & 11 & 690 & & No star formation, 0.5 $h^{-1}$ Mpc box \\
\it{Fiducial runs}
\vspace{-0.1cm}
\\[2mm] \tableline\\
UnderDense$-$\_nH1e4\_fid & $256^3$ & 8 & 22 & 5500 & 10000 &  Underdense box with no massive galaxies, fiducial parameters  \\
UnderDense+\_nH1e4\_fid & $256^3$ & 8 & 22 & 5500 & 10000 & Underdense box with one massive galaxy, fiducial parameters \\
\vspace{0.4cm}
OverDense\_nH1e4\_fid & $256^3$ & 8 & 22 & 5500 & 10000 & Overdense box, fiducial parameters \\
\it{Density threshold study}
\vspace{-0.1cm}
\\[2mm] \tableline\\
OverDense\_nH1e3 & $256^3$ & 8 & 22 & 5500 & 1000 & Lowest density threshold for Pop III star formation \\
OverDense\_nH5e3 & $256^3$ & 8 & 22 & 5500 & 5000 & Low density threshold for Pop III star formation \\ 
\vspace{0.4cm}
OverDense\_nH2e4 & $256^3$ & 8 & 22 & 5500 & 20000 & High density threshold for Pop III star formation \\
\it{Mass resolution \& refinement criteria}
\vspace{-0.1cm}
\\[2mm] \tableline\\
OverDense\_SL7 & $256^3$ & 8 & 22 & 5500 & 10000 & Super-Lagrangian refinement $0.7^\ell$ \\ 
OverDense\_SL5 & $256^3$ & 8 & 22 & 5500 & 10000 & Aggressive super-Lagrangian refinement $0.5^\ell$ \\ 
OverDense\_HiRes & $512^3$ & 7 & 22 & 690 & 10000 & Higher mass resolution\\
OverDense\_HMpc & $256^3$ & 7 & 22 & 690 & 10000 & 0.5 $h^{-1}$ Mpc box \\
OverDense\_HMpc\_HiRes & $256^3$ & 8 & 11 & 690 & 10000 & 0.5 $h^{-1}$ Mpc box, higher spatial resolution \\
OverDense\_HMpc\_SL5 & $256^3$ & 7 & 22 & 690 & 10000 & 0.5 $h^{-1}$ Mpc box, super-Lagrangian refinement $0.5^\ell$ \\
\vspace{0.4cm}
OverDense\_QMpc & $256^3$ & 8 & 5.5 & 86 & 10000 &   0.25 $h^{-1}$ Mpc box \\ 
\it{Alternative physics}
\vspace{-0.1cm}
\\[2mm] \tableline\\
OverDense\_ExtremeSN & $256^3$ & 8 & 22 & 5500 & 10000 &  Extreme PISNe (Section \ref{sec:extremeFB}) \\
OverDense\_ExtremeRad & $256^3$ & 8 & 22 & 5500 & 10000 &  Extreme Pop III radiation field (Section \ref{sec:extremeFB}) \\
OverDense\_LowMass & $256^3$ & 8 & 22 & 5500 & 10000 & Pop III IMF and feedback mirror Pop II (Section \ref{sec:noIII}) \\

\\[2mm] \tableline
\end{tabular}
\end{center}
\vspace{0.3cm}
Column 1.) Name of the run; 2.) Base grid, number of DM particles, number of root cells; 3.) Maximum number of additional levels of refinement; 4.) Minimum cell size at the highest level of refinement in comoving pc; 5.) DM particle mass in $\Msun$; 6.) Minimum H number density for Pop III star formation in $\dim{cm}^{-3}$; 7.) Further description of the run.
\vspace{0.3cm}
\end{table*}

\subsubsection{Density threshold for Pop III star formation}
\label{sec:nHtest}

 \begin{figure}[t]
\vspace{-0.1cm}
\centerline{\epsfxsize3.5truein \epsffile{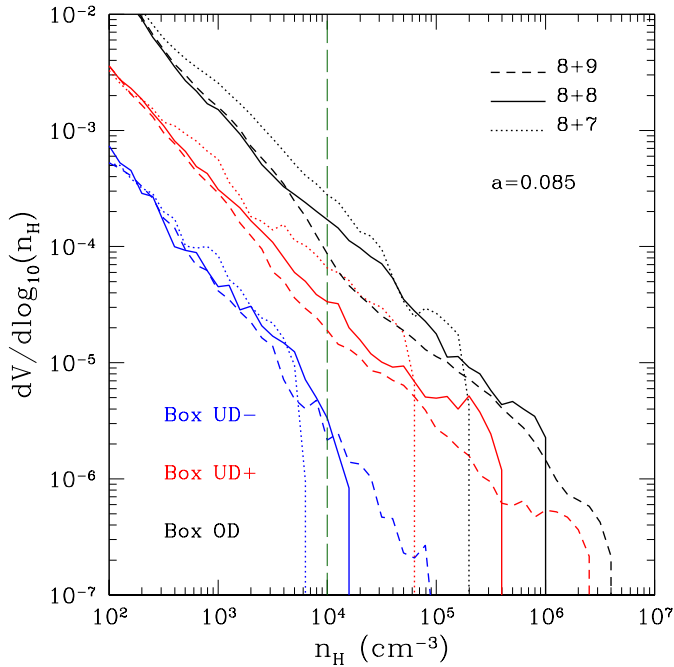}}
\vspace{-0.0cm}
\caption{The distribution function of hydrogen number density for runs without star formation at $a=0.085$ ($z=10.8$). Blue lines are for Box UnderDense$-$, red for Box UnderDense+, black for Box OverDense. Dotted lines are for 8+7 levels of refinement, solid lines for 8+8, short-dashed lines for 8+9. The long-dashed green line represents our chosen density threshold $n_{H, min} = 10000 \dim{cm}^{-3}$. All runs with at least 8 levels of refinement have sufficiently dense gas to form stars by this epoch.}
\vspace{0.3cm}
\label{fig:PDF}
\end{figure}

First, we choose an appropriate value for the density threshold for creating Pop III stars. In Figure \ref{fig:PDF}, we examine the high-density end of the volumetric probability distribution function (PDF) of the hydrogen number density at $a=0.085$ ($z=10.8$). In order to test the properties of the primordial gas from which the first stars form, we ran a special set of simulations with no star formation or chemical enrichment (runs OverDense\_noSF\_8, UnderDense+\_noSF\_8, UnderDense$-$\_noSF\_8, as well as additional versions of each with one more and one fewer maximum level of spatial refinement). Though the total mass of gas in each box is the same, only gas in the most massive halos has collapsed to this density regime, meaning that the PDFs are very sensitive to the number and nature of such halos. The PDFs of Box OverDense and UnderDense+ are offset by a factor of \textasciitilde5 at all densities, while the UnderDense$-$ box is offset from Box UnderDense+ by another factor of \textasciitilde5. This difference is also seen in the maximum density achieved in each box. For our fiducial resolution of 8+8 levels, all three boxes are able to reach a density of at least 10000 $\dim{cm}^{-3}$ by $a=0.085$, giving us enough time to study star formation in every box before our stopping point of $a=0.1$. 

Based on these results, we chose $n_{H, min} = 10000 \dim{cm}^{-3}$ as our fiducial value for the density threshold. In addition to the constraints obtained from the PDF, other considerations went into this selection. A lower value would suffice to meet our proto-cloud collapse criteria, but would result in all Pop III stellar particles forming before cells are maximally refined. Such an outcome is poor practice in hydrodynamic simulations, as subgrid physics is being invoked on scales where the resolution is still good enough to self-consistently capture relevant physical processes. On the other hand, using a higher threshold would allow the maximally refined cells to reach densities beyond the resolving power of the simulation. When such conditions are reached, either further refinement or subgrid physics should already be in use. In addition, using a higher density threshold in our test runs often led to Pop III stars forming in bursts (in the same timestep, in neighboring cells). We do not speculate here whether such bursts are physically plausible or not, but the scales necessary to model this process properly are certainly unresolved in our simulations. We suspect that higher temporal or spatial resolution would reveal that feedback from the first star in a cell would suppress, or at least delay further clustered star formation, as the $H_2$ photo-dissociation timescales due to internal Lyman-Werner feedback from a single $100 \, \Msun$ star within a given star-forming clump are typically shorter than the clump's free-fall timescale \citep{safranek_etal12}. 

To determine the ultimate effect of the density threshold on Pop III star formation, additional runs were performed with $n_{H, min} = 1000$, $5000$, and $20000 \dim{cm}^{-3}$ using the Box OverDense initial conditions. Varying this threshold by a factor of 20 changes the scale factor at which the first star forms only from $a=0.0463$ to $a=0.0483$, or from redshift $z=20.6$ to $z=19.7$. In the $n_{H, min} = 1000 \dim{cm}^{-3}$ run, the density threshold is reached at a lower level of refinement from the other cases considered, allowing the first star to form sooner. The variation in the other three runs is only from $a=0.0480$ to $a=0.0483$. Such marginal differences demonstrate that for a given set of initial conditions, our actual density threshold criterion for the formation of Pop III stars has little effect on when and where they form. In each of these runs, only a single Pop III star formed in each box before $a=0.05$, and the total number of Pop III stars varied between 4 and 5 at $a=0.055$. Based on these tests, we have determined that the total number of Pop III stars formed had little correlation with the density threshold within the considered range.

\begin{figure}[t]
\vspace{-0.2cm}
\centerline{\epsfxsize3.5truein \epsffile{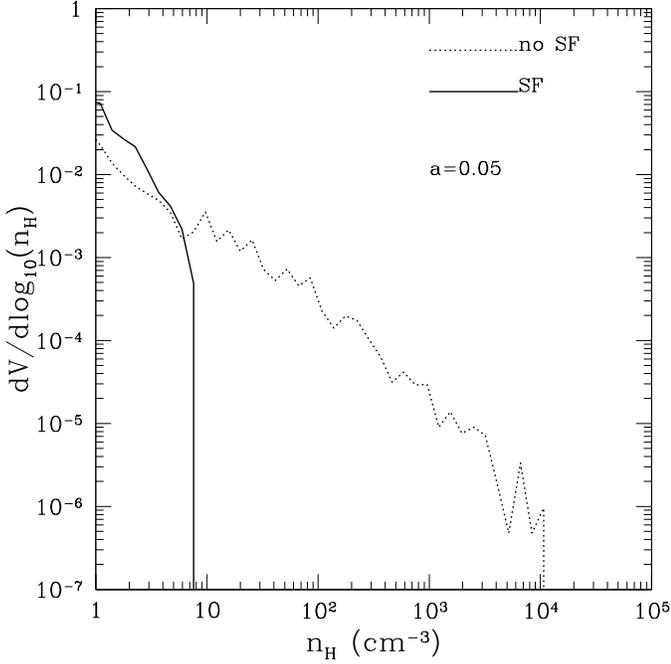}}
\vspace{-0.0cm}
\caption{ PDF for most massive galaxy in Box OverDense at $a=0.05$ ($z=19$) with star formation (solid, run OverDense\_nH1e4\_fid) and without star formation (dotted, run OverDense\_noSF\_8). The Pop III star that recently formed at the center of this galaxy has temporarily depleted it of the dense gas needed to continue star formation. }
\vspace{0.3cm}
\label{fig:PDF2}
\end{figure}

After the first star forms in a given halo, the gas density can be significantly reduced near the center, quenching further star formation. Figure \ref{fig:PDF2} demonstrates this effect in run OverDense\_nH1e4\_fid. The PDF of this galaxy is depleted at high density 10 Myr after a PISN explosion. The corresponding galaxy from the run without star formation, OverDense\_noSF\_8, is also shown for reference. While the galaxy in OverDense\_noSF\_8 contains some dense gas above $n_H = 10 \dim{cm}^{-3}$, it is depleted in our fiducial run. Since this density is nowhere near any $n_{H, min}$ that we have considered in our tests, we can conclude that Pop III stars will not form in quick succession in this halo.

\ 

\subsubsection{Molecular Fraction}
\label{sec:molfrac}

\begin{figure}[t]
\vspace{-0.2cm}
\centerline{\epsfxsize3.5truein \epsffile{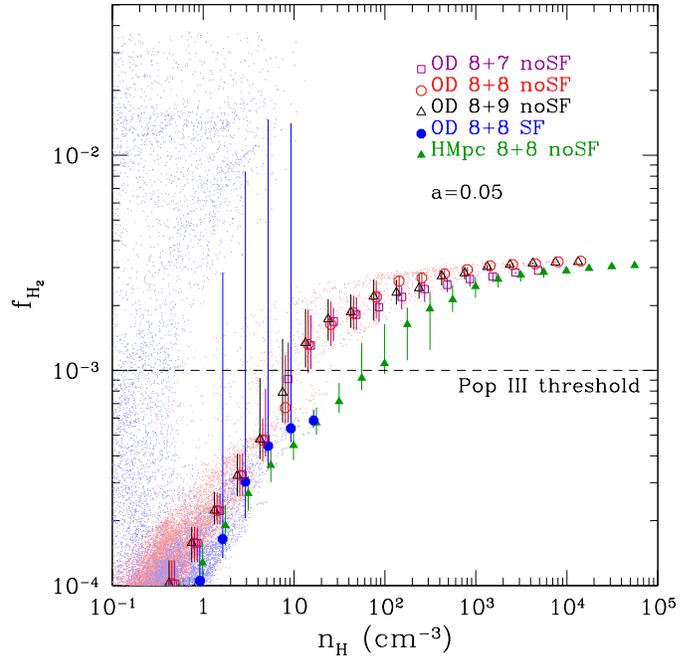}}
\vspace{-0.0cm}
\caption{Molecular fraction of hydrogen vs number density for three Box OverDense runs using different resolution without star formation at $a=0.05$ ($z=19$, runs OverDense\_noSF\_9, OverDense\_noSF\_8, and OverDense\_noSF\_7 are black triangles, red circles, and purple squares, respectively). The median $H_2$ fraction in each density bin is indicated by a triangle, while the error bars show 25th and 75th percentile levels. At this epoch, when the first stars would normally be forming, our fiducial resolution of 8 levels has the same molecular fraction as if we were using one more or one fewer level of refinement, and the points actually lie directly on top of each other for $n_H < 10 \dim{cm}^{-3}$. Run OverDense\_noSF\_HMpc at $a=0.05$ (green filled triangles) has lower values and wider spread of $H_2$ fraction for $n_H < 10^2 \dim{cm}^{-3}$ but converges with the other runs at higher densities, demonstrating a lack of dependence on mass resolution. Also shown is run OverDense\_nH1e4\_fid where star formation has already taken place by $a=0.05$ (blue filled circles). Since the gas has been enriched by a PISN, molecular gas can form at much lower densities. Light red and light blue points trace out the $H_2$ fraction in every single cell for run OverDense\_noSF\_8 and OverDense\_nH1e4\_fid, respectively.}
\vspace{0.3cm}
\label{fig:fH2nH}
\end{figure}

Another component of the Pop III star formation criterion is the requirement of a minimal fraction of $H_2$ in the host cell. To determine what value of the $H_2$ threshold makes sense in the context of these simulations, we examine the molecular fraction of hydrogen as a function of density in the runs without star formation at the epoch ($z\approx20$) when gas is beginning to reach densities close to $n_{H, min}$. Figure \ref{fig:fH2nH} shows that the molecular fraction in primordial gas generally increases with density, but saturates above $n_H \approx 10 \dim{cm}^{-3}$. The saturation value of $f_{H_2}$ grows slowly with time in the absence of star formation, and does not appear to depend significantly on spatial or mass resolution. Our fiducial choice of $10^{-3}$ for the minimal $H_2$ fraction does not exclude dense gas from forming stars in any runs, as long as little Lyman-Werner radiation is present.

\subsubsection{Super-Lagrangian Refinement}
\label{sec:SLR}
Since we have demonstrated resolution dependence for the maximum density of gas within a given galaxy, it is expected that refinement criteria could play a role in controlling when gas in the simulation first reaches the $n_{H, min}$ threshold. To test this, in some of our runs we employ super-Lagrangian (SL) refinement criteria. This approach dictates that the refinement threshold between subsequent levels is lowered by a constant factor, granting a more effective zoom-in on the densest regions at earlier times. The refinement criteria in a cell can be written as $2 \times m_{\, DM} \times \frac{\Omega_m}{\Omega_{DM}} \times X^{\ell} \times 0.8$ for the dark matter mass, and $0.3 \times m_{\, DM} \times \frac{\Omega_m}{\Omega_{DM}} \times X^{\ell} \times 0.8$ for the gas mass, where $\ell$ is the level of the cell which is to be refined. In this formalism $X=1$ implies Lagrangian refinement as described at the beginning of Section \ref{sec:sims}. We have tried runs with very aggressive SL refinement ($X=0.5$, run OverDense\_SL5) and less aggressive refinement ($X=0.7$, run OverDense\_SL7). Running these simulations in Box OverDense with $n_{H, min}=10000 \dim{cm}^{-3}$, we found that the epoch at which Pop III stars first appear is pushed back from $a=0.0478$ to $a=0.0456$ with $X=0.7$, and to $a=0.0435$ with $X=0.5$. This demonstrates that the use of SL refinement is an important numerical tool for exploring the earliest epoch of star formation in a given simulation box. However, using SL refinement produces an enormous number of high-level cells: at $a=0.05$, run OverDense\_SL5 has a factor of 4000 more maximally refined cells than run OverDense\_nH1e4\_fid. This drastic difference makes the SL simulations prohibitively expensive soon after the first stars form. 

Therefore, we use these SL runs to study Pop III star formation at the earliest possible epochs, when the mass of the halos that hosted them was low enough for PISNe to have their maximal effect. 

\subsubsection{Increased Mass Resolution}
\label{sec:512methods}
We test the effects of mass resolution by setting up one run with $512^3$ initial grid, giving a DM particle mass of $690 \, \Msun$. We use 7 additional levels of refinement, therefore granting us the same maximum spatial resolution as in the fiducial $256^3$ run. Having consistency in spatial resolution allows us to test the effects of mass resolution alone. All other numerical parameters are kept consistent with run OverDense\_nH1e4\_fid. 

The increased mass resolution has several immediate implications. Since we now resolve halos of mass $10^6 \, \Msun$ with over 1000 particles, we can better probe the regime where the very first Pop III stars are expected to collapse in proto-galactic 'minihalos'. Indeed, the epoch of formation of the first star in the box becomes $a=0.0427$ in a halo of $1.5 \times 10^6 \, \Msun$ (compared to $a=0.0481$ and $M_h=7.5\times10^6\, \Msun$ in run OverDense\_nH1e4\_fid). The higher mass resolution effectively means that there is more power on the small scales responsible for the growth of halos in this mass regime. Due to the high computational cost of this run, we have only advanced it to $a=0.055$. 

The effect of increasing mass resolution is further explored through runs using the H and Q boxes of 0.5 $h^{-1}$ Mpc and 0.25 $h^{-1}$ Mpc in size. Applying the $256^3$ base grid to these boxes gives us a DM particle mass of $690 \, \Msun$ and $86 \, \Msun$, respectively. We find that the H box (run OverDense\_HMpc\_HiRes) produces a Pop III star by $a=0.0456$ in a halo of mass $1.5 \times 10^6 \, \Msun$, while the Q Mpc box (run OverDense\_QMpc) does not produce one until $a=0.0473$. Using SL refinement in the HMpc box (run OverDense\_HMpc\_SL) allows us to see a Pop III star forming in a $8 \times 10^5 \, \Msun $ halo. 

\begin{figure}[t]
\vspace{-0.2cm}
\centerline{\epsfxsize3.5truein \epsffile{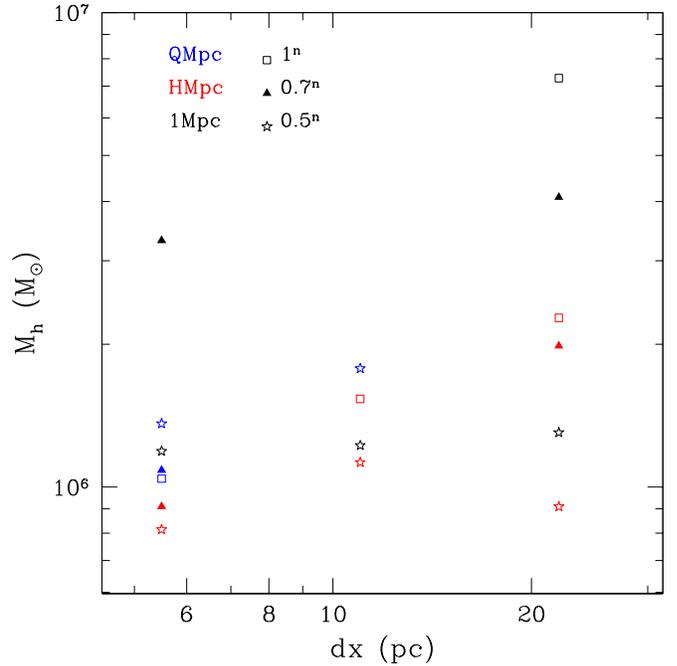}}
\vspace{-0.0cm}
\caption{ The least massive galaxy to host a star in various runs vs. the minimum comoving cell size employed in the run. The colors indicate the simulation box size: 1 $h^{-1}$ Mpc (black), 0.5 $h^{-1}$ Mpc (red), or 0.25 $h^{-1}$ Mpc (blue). The shape of points indicates the refinement criterion employed in the box, with open squares for Lagrangian refinement, filled triangles for $0.7^\ell$ SL refinement, and open five-pointed stars for $0.5^\ell$ SL refinement.}
\vspace{0.3cm}
  \label{fig:MOFH}
\end{figure}

The earlier formation epochs and lower mass of halos hosting the first stars in the H and Q Mpc boxes, compared to the fiducial 1 $h^{-1}$ Mpc runs, show that it is crucial to have high enough mass resolution to capture Pop III star formation in halos close to $10^6 \, \Msun$. It has been previously shown that halos less massive than this threshold will not achieve significant enough $H_2$ abundances to trigger Pop III star formation at earlier epochs \citep{yoshida_etal03}. The further significance of this mass range will be explained in our Results section. Figure \ref{fig:MOFH} shows explicitly how varying refinement criteria, spatial resolution, and mass resolution affected the lowest possible mass for a star-forming galaxy.

\subsubsection{Low Mass Pop III IMF}
\label{sec:noIII}
We present one simulation, run OverDense\_LowMass, which does not rely on a top-heavy IMF for Pop III stars. The conditions for Pop III star formation in this run are similar to our fiducial top-heavy recipe in that we use the same threshold $n_{H, min}$ to determine which cells are allowed to form stars. However, the stellar particle masses are drawn from the same IMF as for Pop II stars. This run explores the possibility that Pop III stars were ordinary low-mass objects. Whenever the density in a given cell exceeds the threshold, the star formation rate is determined according to the following relation:
\begin{equation}
\dot{\rho_*} = \frac{\epsilon_{ff}}{\tau_{SF}} \rho_{gas},
\end{equation}
where $\rho_{gas}$ is the mass density of all gas in the cell. This relation is similar to equation \ref{eq:SFC}, but does not explicitly use molecular hydrogen. This modification is necessary because primordial gas can reach densities above our star formation threshold, but cannot become fully molecular without the presence of dust.

\subsubsection{Extreme Pop III Feedback}
\label{sec:extremeFB}

To isolate the relative impacts of the feedback effects, we ran toy simulations using exaggerated values for the PISN energy and ionizing photon yield. In one run, called OverDense\_ExtremeSN, PISNe released $270 \times 10^{51}$ erg of thermal energy, a factor of 10 larger than in all other runs. The extreme ionizing simulation OverDense\_ExtremeRad had instead an additional factor of 10 boost in the ionizing photon flux of Pop III stars, giving the $100 \, \Msun$ and $170 \, \Msun$ stars 388,000 and 345,000 photons per lifetime, respectively. While these models are too strong to be consistent with any published results, using them allows us to explore the most extreme effects of Pop III feedback. 

\section{Results}
\label{sec:results}

\begin{figure}[t]
\vspace{-0.2cm}
\centerline{\epsfxsize3.5truein \epsffile{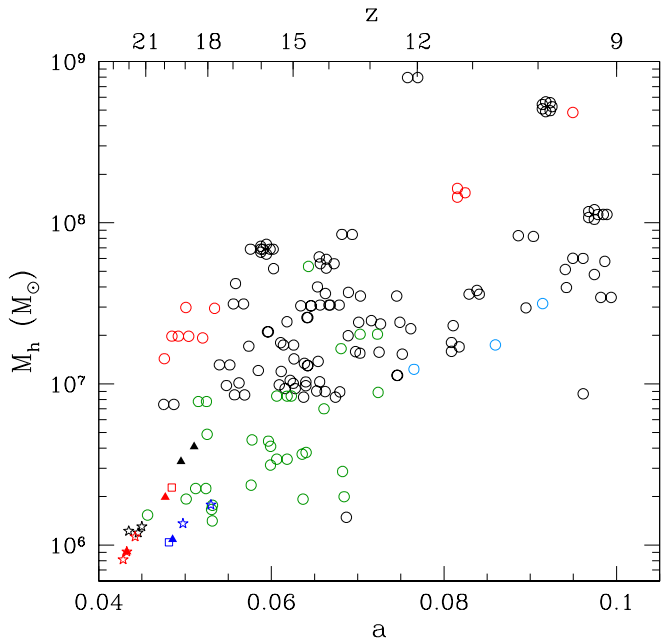}}
\vspace{-0.1cm}
\caption{Each Pop III star's host halo mass at the time of formation vs the scale factor at which the star formed for run UnderDense$-$\_nH1e4\_fid (blue), run UnderDense+\_nH1e4\_fid (red), run OverDense\_nH1e4\_fid (black), and run OverDense\_HMpc\_HiRes, which did not go past $a=0.075$ (green). Additional points for the halos hosting the first stars from each of the runs used for Figure \ref{fig:MOFH} are also included, with the same color and shape scheme. In our 1 $ h^{-1}$ Mpc runs, Pop III star formation happens almost exclusively in halos between $10^7 \, \Msun$ and $10^8 \, \Msun$. The additional mass resolution in run OverDense\_HMpc\_HiRes makes it possible to see that the first Pop III stars form in halos between $10^6 \, \Msun$ and $10^7 \, \Msun$. The average mass of Pop III star-forming halos increases slightly with time. When multiple Pop III stars form within a galaxy in a very short time interval, points on the plot are grouped into a clustered shape.}
\vspace{0.3cm}
\label{fig:Mha}
\end{figure}

Pop III stars in our simulations begin to form in halos of mass $M_h \gtrsim 10^6 \, \Msun$ starting at $a\approx0.045$ ($z\approx21.2$) in accordance with expectations from prior work \citep{yoshida_etal03}. Figure \ref{fig:Mha} shows the mass of host halo in which each Pop III star formed. Pop II stars begin forming in most of these halos shortly thereafter, but are not shown in this plot. In the 1 $ h^{-1}$ Mpc runs, the halo mass for first star formation is close to $10^7 \, \Msun$. This mass is an order of magnitude larger than that of the halos hosting the first stars in the simulations of \citet{wise_etal12} and \citet{greif_etal11}, and those preferred by theoretical considerations \citep{tegmark_etal97}. Consequently, those authors also find an earlier epoch for the formation of the first stars. Given that the extra mass resolution granted by the H Mpc box allows us to see star formation in $10^6 \, \Msun$ halos, we infer that our fiducial 1 $h^{-1}$ Mpc runs are not properly resolving the very first star-forming minihalos. Rather, they are more generally simulating Pop III star formation in an early population of galaxies. The fraction of star-forming halos in run OverDense\_nH1e4\_fid at $a=0.07$ ($z=13.3$) was only 1\% in the mass range $10^6 \, \Msun < M_h < 10^7 \, \Msun$, but it reached 65\% for $M_h > 10^7 \, \Msun$. In run OverDense\_HMpc\_HiRes, these numbers increase significantly to 15\% for $10^6\, \Msun < M_h < 10^7 \, \Msun$ and 100\% for $M_h > 10^7 \, \Msun$. In addition to the resolution effects, the suppression of star formation in the $10^6$ to $10^7$ $\Msun$ range is also plausible in the regime of a moderate Lyman-Werner background (e.g. \citealt{machacek_etal01, oshea_norman08, safranek_etal12}).

It is worth noting that the ratio of star-forming galaxies in the range $10^7 \, \Msun < M_h < 10^8 \, \Msun$ falls off gradually with time in run OverDense\_nH1e4\_fid. It changes from 65\% at $a=0.065$ to 20\% at $a=0.1$, suggesting that halo mass alone is not a good proxy for determining whether a galaxy can achieve the high density required for our Pop III star formation criteria. One potential cause of the change is the decreased physical spatial resolution at later epochs, but according to our study of the gas in the first star-forming galaxy shown in Figure \ref{fig:fH2nH}, the factor-of-two difference in spatial resolution achieved by using one fewer level of refinement does not preclude gas from reaching the fiducial $n_{H,min}=10^4 \dim{cm}^{-3}$ threshold. The difference is more likely to be rooted in the evolution of physical density in halos of a given mass. For halos between $10^7$ and $10^8 \, \Msun$, the average matter density within the virial radius changes from $2.5 \times 10^{-2} \, \Msun \dim{pc}^{-3}$ at $a=0.065$ to $6.8 \times 10^{-3} \, \Msun \dim{pc}^{-3}$ at $a=0.1$, due to the expansion of the universe. Even the density within the central 100 pc of these halos changes from $0.73 \, \Msun \dim{pc}^{-3}$ to $0.34 \, \Msun \dim{pc}^{-3}$ between the same two epochs.

Very few Pop III stars formed in halos with $M_h > 10^8\, \Msun$, because such halos have already been enriched to metallicities above $\log_{10} Z/Z_\odot = -3.5$, allowing for normal star formation to commence. In many cases, halos with $M_h > 3 \times 10^7 \, \Msun$ had earlier formed one or more $100 \, \Msun$ Pop III stars, which shut off star formation temporarily but did not enrich the galactic gas, allowing it to remain pristine and continue forming Pop III stars. Another major exception occurs in run OverDense\_nH1e4\_fid in a galaxy that has already formed a significant number of Pop II stars that have in turn enriched the ISM terminating further Pop III star formation. However at $a=0.095$, this galaxy undergoes a major merger with another massive halo, causing low-metallicity gas in the outer part of the galaxy to collapse, thereby triggering a burst of Pop III star formation which appears as a cluster of points with $M_h = 5 \times 10^8$ on Figure \ref{fig:Mha}. All of these stars form with metallicities around the critical $\log_{10} Z/Z_\odot =-3.5$ value, suggesting that their existence is sensitive to the value of this threshold and therefore should not be treated as a general result.

\subsection{Effect of Pop III stars on their host galaxies}
 \begin{figure}[t]
\vspace{-0.2cm}
\centerline{\epsfxsize3.5truein \epsffile{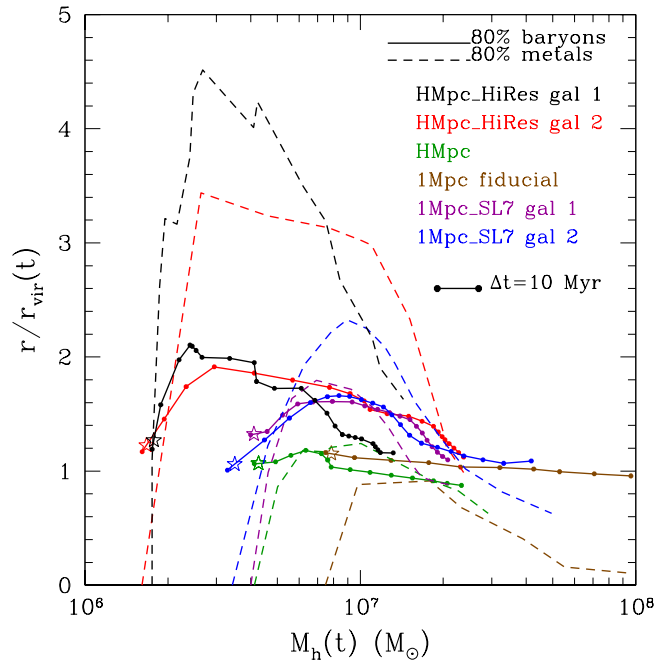}}
\vspace{-0.1cm}
\caption{ Though metals from PISNe
are ejected past the virial radius, they do not stay there
for long. Plotted here are the radii enclosing 80\% of the
metals produced in the galaxy, as well as
the radii where the enclosed mass of baryons divided
by the virial mass equals 80\% of the universal baryon
fraction. Each line represents a galaxy as it evolves in
time, beginning at the epoch when the first star forms. Within 50 Myr of the PISN event (denoted by five-pointed stars),
most of the gas and metals have begun to recollapse, or are at least enclosed within the virial radius once again.
The maximum extent of metal propagation is strongly regulated by
galaxy mass. The distance between two points on each line corresponds to 10 Myr.}
\vspace{0.3cm}
\label{fig:Mh_ratios}
\end{figure}

The strong ionizing flux of Pop III stars and enormous energy injections from PISNe have been shown in previous work to significantly alter the ISM of their host galaxies. Here we explore such effects during time when Pop III stars are the dominant drivers of feedback. 

In Figure \ref{fig:Mh_ratios} we show that there is a significant variation in the potential effect of Pop III stars on their host galaxies depending on the galaxy mass. In halos with $M_h < 3 \times 10^6 \, \Msun$, Pop III stars can temporarily evacuate the gas from the galaxy, and the metals from PISNe are ejected past the virial radius into the intergalactic medium (IGM). On the other hand in halos with $M_h > 3 \times 10^6 \, \Msun$, the metals are confined within the virial radius, and there is little movement of baryons beyond the virial radius. 

\begin{figure}[t]
\vspace{-0.2cm}
\centerline{\epsfxsize3.5truein \epsffile{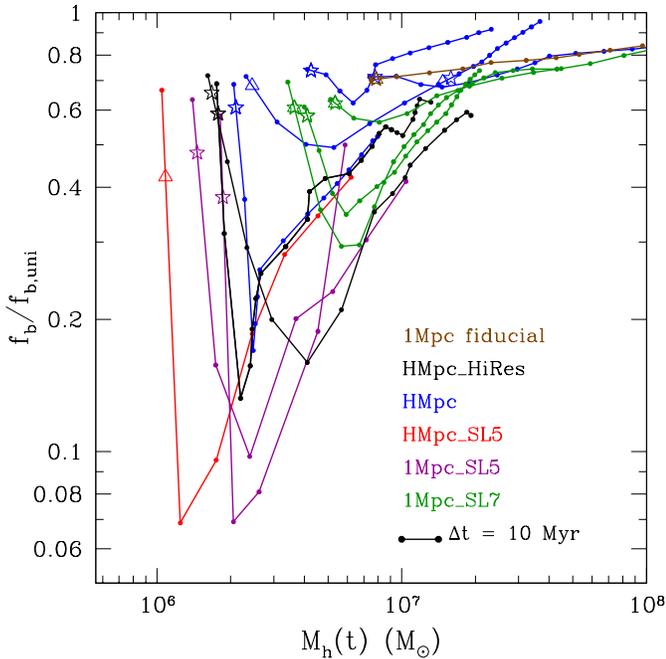}}
\vspace{-0.1cm}
\caption{ The evolution of the baryon fraction vs. the evolving halo mass. Each line traces a galaxy from the time of formation of the first star. The baryon fraction is computed within the virial radius, and is normalized by the universal value $f_{b,uni}$. While supernovae initially cause a depletion of baryons in galaxies of $M_h < 3 \times 10^6 \, \Msun$, this depletion is only temporary. In galaxies of $M_h > 3 \times 10^6 \, \Msun$, there is no strong evidence that PISNe are able to deplete baryon fractions. Notation is the same as in Figure \ref{fig:Mh_ratios}, with the addition of triangles to represent SN II produced by $100 \, \Msun$ Pop III stars.}
\vspace{1.0cm}
\label{fig:Mh_fbar}
\end{figure}

This divide can also be seen in Figure \ref{fig:Mh_fbar}, which shows the baryon fraction computed within the virial radius for thirteen halos taken from a variety of runs. Again, a noticeable threshold at $M_h = 3 \times 10^6 \, \Msun$ distinguishes galaxies that have their gas evacuated by PISNe from those that do not. Galaxies less massive than this threshold typically have their gas content plummet by at least a factor of two within 10-30 Myr after the PISN, with the least massive ones falling below 10\% of the universal baryon fraction. In contrast, more massive galaxies lose a much smaller percentage of their gas and end up with baryon fractions in excess of their pre-explosion values within \textasciitilde100 Myr. This dividing line between "low-mass" and "high-mass" galaxies is therefore a logical choice for parameter that distinguishes different regimes of Pop III feedback. We examine these two regimes separately below. 

\subsubsection{Dependence on halo mass}
\label{sec:depmass}
To further understand how halo mass can determine the effectiveness of Pop III stellar feedback, we examine the structural evolution of several galaxies in different mass regimes. 

First, we examine a relatively low-mass galaxy from run OverDense\_HMpc\_HiRes, which is shown by the black line that extends to $M \approx 10^7 \, \Msun$ in Figures \ref{fig:Mh_ratios} and \ref{fig:Mh_fbar}. The first star (of $170 \, \Msun$) forms when the mass of the halo is $1.7 \times 10^6 \, \Msun$. Within 20 Myr of its formation, the PISN has blasted metals out beyond 1 kpc from the galactic center (or $4 R_{vir}$ at this epoch), and the baryon fraction has dipped to $f_b / f_{b,uni} \approx 0.15$. However, soon after this point the baryon fraction begins to grow again, and the virial radius increases enough to enclose a larger fraction of the expelled gas and metals. 

At $t=139$ Myr after the first PISN (the halo mass is now $9\times 10^6 \, \Msun$), the galaxy has regained \textasciitilde37\% of the PISNe metals. The baryon fraction is over half of the universal value. It will still take more time for this galaxy to completely recover from the explosion, but there is considerable evidence from Figure \ref{fig:Mh_ratios} that metals and baryons in general are flowing in rather than out of the galaxy. Another sign of recovery is that Pop II star formation has commenced within the galaxy, as it now contains 4 Pop II stellar particles (which still contribute little to the metal budget).

\begin{figure}[t]
\vspace{-0.2cm}
\centerline{\epsfxsize3.5truein \epsffile{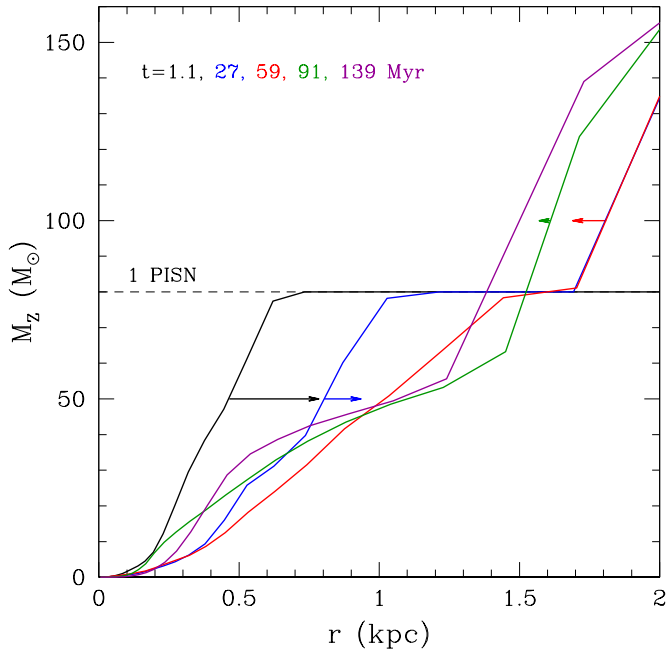}}
\vspace{-0.1cm}
\caption{ The ejecta of PISNe is traced via examining the enclosed mass of metals as a function of galactocentric distance. Lines of different color correspond to 1, 27, 59, 91, and 139 Myr after the first PISN. A second PISN happens 15 Myr after the first in a nearby galaxy. Approximately 60 Myr after the first PISN, more metals are flowing into the galaxy than outwards, as the metal-rich ejecta have mixed with primordial gas accreting onto the galaxy. Arrows show the direction of metal movement at each epoch. The length of each arrow corresponds to the distance traversed by the metals in a 20 Myr interval. The y-axis positions of the arrows show the mass of metals at each epoch used to compute the rate of propagation. This galaxy is depicted by the black line in Figures \ref{fig:Mh_ratios} and \ref{fig:Mh_fbar}.}

\vspace{0.3cm}
\label{fig:FollowMetals}
\end{figure}

Figure \ref{fig:FollowMetals} follows the radial distribution of metals in this galaxy from the time of the first PISN until 139 Myr after it has exploded. The Pop II stars have contributed less than 1 $\Msun$ to the metal budget, so essentially all of the metals shown here are products of the first PISN, and of a second PISN which happens 15 Myr after the first in a neighboring halo at a distance of approximately 2 kpc. The mass of this galaxy has increased by a factor of \textasciitilde5 between $a=0.0508$ when the star first formed and $a=0.0726$ at the final snapshot considered.

While the PISNe do clearly cause baryon depletion and suppress star formation in galaxies such as the one presented here, the rate of growth of these galaxies is high enough that a mixture of ejecta and fresh primordial gas fall in to restore the baryon fraction to at least 50\% of the universal fraction within \textasciitilde150 Myr. This replenishment results from a combination of actual re-accretion of ejected material, accretion of new primordial baryons from filaments, and rapid growth of galaxy virial radius ("gobbling up" of ejecta).

The eventual fate of this low-mass galaxy, and of many such minihalos which were significantly affected by the first PISNe, is to merge with a more massive companion prior to the completion of the metal re-accretion process. The resultant galaxy will ultimately have a baryon fraction close to the universal value, and will contain enough of the PISN ejecta from the progenitors to form Pop II stars. In some cases, we observed that halos in this mass range sustained more long-term damage from PISN, and their baryon fraction stayed below 50\% by the end of our simulation, as late as 200 Myr after the explosion. This scenario played out in relatively isolated environments with slow filamentary accretion. Galaxies that underwent such long-term disruption by PISNe had their virial mass increase at an average rate of $0.04 \, \Msunyr$ for 100 Myr after the explosion, while all other galaxies that hosted Pop III stars grew at rates ranging from $0.04 \, \Msunyr$ to $0.8 \, \Msunyr$. 

In run OverDense\_HMpc\_HiRes, which effectively resolved galaxies in the minihalo regime, 21\% of PISNe occurred in underdense environments where the metals were permanently ejected from the host galaxy. Another 25\% of PISNe happened in galaxies where the host merged with a separate galaxy prior to the complete gobbling of metals. The remaining 54\% of PISNe happened in galaxies where metals were effectively gobbled up by the end of the simulation. These findings suggest that 20-45\% of the metals from Pop III supernova ejecta can be observed in the IGM at $z\approx10$.

%However, these cases were the exceptions and ultimately Pop III stars have little direct impact on the IGM.  FO
%as evidenced by the relatively low mass of these galaxies at later times FO. 

Next, we study a galaxy from run OverDense\_nH1e4\_fid that was the first to form a Pop III star, which also happens to be a PISN progenitor. This galaxy is represented by the brown lines in Figures \ref{fig:Mh_ratios} and \ref{fig:Mh_fbar}. The first star forms when the mass of the halo is $7.5 \times 10^6 \, \Msun$. About 30 Myr later, 80\% of the metals generated in the PISN have propagated as far as 420 parsecs from the core. The injection of metals by the PISN is enough to bring the gas metallicity in some cells hundreds of parsecs away from the galactic center to be as high as $\log_{10} Z/Z_\odot = -1$. After another 50 Myr, the effects of the outflow have subdued. Not only are 80\% of generated metals now entirely confined to the innermost 120 parsecs, but the metals have diffused, and the maximum metallicity has decreased by 1 dex. This suggests that the inflow of new primordial gas is playing a greater role in the evolution of the galaxy than the outflow generated by the PISN. The majority of metals produced by PISN do not escape into intergalactic space.

We emphasize that some galaxies in this mass range should have hosted Pop III stars at earlier times than those resolved by our simulations. The apparent ineffectiveness of Pop III feedback demonstrated here shows that simulations which do not resolve halos with $M_h < 3 \times 10^6 \, \Msun$ are missing a portion of galactic evolution. This omission could mean that Pop III stars should self-terminate at earlier times, hence decreasing their contribution to the cosmic ionizing background. On the other hand, the expulsion of baryons from low-mass halos leads to suppression of Pop II star formation, which implies we may also overestimate the Pop II rates. The balance of these effects will be explored further in Paper II. 

\begin{figure}[t]
\vspace{-0.2cm}
\centerline{\epsfxsize3.5truein \epsffile{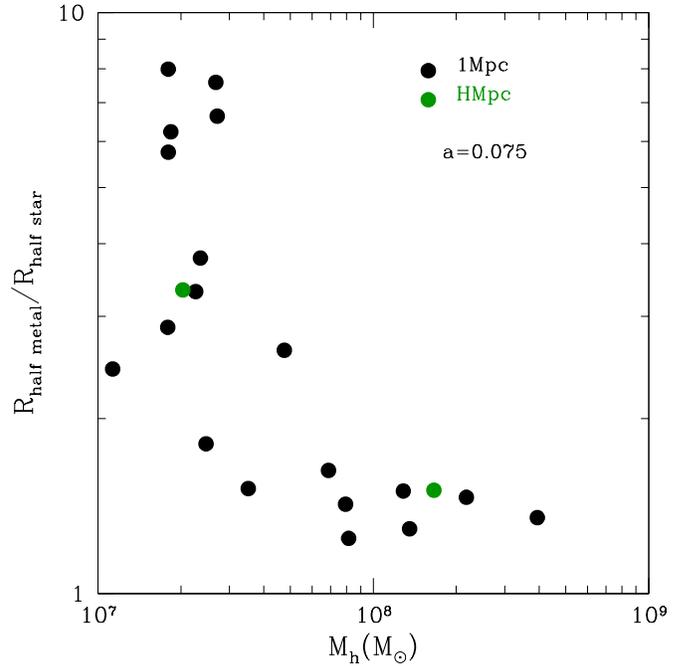}}
\vspace{-0.1cm}
\caption{ The ratio of gas metal half-mass radius to stellar half-mass radius vs halo mass for galaxies which have had at least one PISN, at $a=0.075$ ($z=12.3$) in runs OverDense\_nH1e4\_fid (black) and OverDense\_HMpc\_HiRes (green). Metals propagate further relative to the stellar cores in galaxies of lower mass.}
\vspace{0.3cm}
\label{fig:HalfRadii}
\end{figure}

Figure \ref{fig:HalfRadii} shows how far metals propagate in galaxies relative to the stellar cores. The "gas metal half-mass radius" is calculated as the more familiar stellar half-mass radius, but tracing the total mass of metals in the gas phase instead of stellar mass. In general, the metals are always able to propagate out beyond the stellar cores, but the extent depends strongly on galaxy mass. For a homogeneous comparison, and potential future probes by observation, we plot all star-forming galaxies from a single epoch, $a=0.075$ ($z=12.3$). In galaxies with $M_h \geq 10^8 \, \Msun$, the metals remain within a factor of 2 of the stellar radius. In less massive galaxies, metals are able to propagate further, sometimes by as much as a factor of 10, owing to the lower potential wells of these galaxies. Nonetheless, considering that the stellar half-mass radii of all our galaxies range 5-30 pc, the location of the bulk of metals is still limited to only the innermost regions of galaxies.

\subsection{Effects of the uncertainty in Pop III feedback and IMF}
\label{sec:ExResults}

With our additional runs, we can check if the relatively inefficient feedback is due to the specific fiducial parameters that we adopted. However, even with the extreme Pop III feedback prescriptions described in Section \ref{sec:extremeFB}, we find that the baryon fraction within the virial radius is never significantly depleted. At $a=0.05$ ($z=19$), 10 Myr after the PISN explosion in the first star-forming galaxy, we find $f_b = 10.3 \%$ and $5.8 \%$ in runs OverDense\_ExtremeSN and OverDense\_ExtremeRad, compared to $11.5\%$ in the fiducial run. In the case of the extremely energetic PISN, this is a relatively small resulting difference for a 10-fold increase in the thermal energy and ionizing radiation output. The mass of the galaxy at this epoch is $9 \times 10^6 \, \Msun$, which we have shown to be large enough to withstand standard Pop III feedback. On the other hand, the difference is more pronounced in the case of extreme ionizing feedback, indicating that the disruptive efficacy of supernovae is significantly increased when it explodes in a region where neutral hydrogen has been more effectively ionized and dispersed by radiative feedback.

The effect of both types of extreme feedback on the propagation of metals is stronger. Metals tend to be blown out of galaxies anisotropically, often extending outwards in directions orthogonal to filaments, into lower density regions. The galactocentric radius that encloses 80\% of the metals formed in the PISN stretches out to $1.7 R_{vir}$ 20 Myr after the explosion in both run OverDense\_ExtremeSN and run OverDense\_ExtremeRad, compared to just $0.91 R_{vir}$ in the fiducial run. These metals do not fully escape the gravitational pull of the galaxy, however, and 90 Myr after the explosion, the galaxies from both extreme feedback runs contain 80\% of the metals from the first PISN within the virial radius (in the fiducial run, they are contained within just $0.15 R_{vir}$). 

Though we increased the feedback effects by a factor of 10, only modest and temporary differences were observed between the runs. Such inefficiency of feedback demonstrates the weak coupling of thermal energy from PISNe to the ISM at the densities and temperatures resolved by our simulations, as almost any amount of energy can be quickly radiated away. This can be seen when considering typical cooling times in the ISM, $\tau_{cool} = k_b T / \Lambda n \approx 3000 (T / 10^4 \dim{K}) ( 1 \dim{cm}^{-3} / n) \dim{yr}$, for $\Lambda = 10^{-23} \dim{erg}\, \dim{s}^{-1}\, \dim{cm}^{-3}$ \citep{hopkins_etal11}. The cooling time of the dense, filamentary gas surrounding the supernova remnant ($n=10\dim{cm}^{-3}$, $T=10^4\dim{K}$) is just \textasciitilde300 yr, which is comparable to a typical timestep in our simulations (\textasciitilde500 yr). This dense gas mixes with the shock-heated supernova remnant, allowing the entire region to return to the ambient temperature of the ISM within a few Myr.

The impact of extreme feedback is more pronounced in the IGM, particularly in run OverDense\_ExtremeRad. The mass fraction of ionized gas between 1-3 kpc from the galactic center is enhanced by a factor of \textasciitilde200, compared to the fiducial run, even 40 Myr after the PISN. Within the same distance range, the IGM temperature is a factor of \textasciitilde10 higher at this epoch. The relatively hot and ionized IGM in turn could affect accretion rates onto galaxies at later times.

The effect of making PISNe ten times more powerful in the H Mpc box was more drastic, as this box sampled lower mass galaxies. The baryon fraction in the first galaxy dropped below $10^{-5}$ after the first PISN, compared to 1.7\% in the standard run OverDense\_nH1e4\_fid. The radii enclosing 80\% of the baryons and metals are twice as large as in the standard run, demonstrating that the added energy in this extreme run coupled with the ISM more efficiently. Even with the standard feedback prescription we would expect a strong blowout in a halo of this mass ($2.7 \times 10^6 \, \Msun$ at this epoch). However, in the fiducial run this galaxy ultimately gobbled up most of the ejected metals. On the other hand, the extreme PISN energy ($270 \times 10^{51} \dim{erg}$) is able to completely destroy the high-density gas clouds needed for star formation, prevent re-accumulation of dense gas from filaments, and cause the metal ejecta to travel far enough into the IGM where they may never fall back onto the galaxy.

These tests indicate that given enough energy input, the host halos of Pop III stars can become completely devoid of gas for cosmologically-significant intervals of time, particularly when they are below the mass threshold \textasciitilde$3\times 10^6 \, \Msun$. However, for the feedback parameters currently considered realistic (our fiducial runs), the feedback of the first stars is limited as illustrated in Figures \ref{fig:Mh_ratios} and \ref{fig:Mh_fbar}.

In the run with low masses of Pop III stars (OverDense\_LowMass), without any PISN, metal transport is extremely ineffective. At $a=0.055$, 80\% of the metals that have been generated by stars in the first star-forming galaxy are confined within 75 pc of the galactic center, compared to 420 pc in the fiducial run. This test demonstrates that if Pop III stars did not have a top-heavy IMF, their contribution to enriching the IGM would be further marginalized. These results agree qualitatively with the work of \citet{ritter_etal12}, who argued that filamentary accretion was never significantly disrupted if Pop III stars had low or moderate characteristic masses and exploded in type II supernovae.

\section{Discussion and Conclusions}

We have presented the results of simulations that implemented primordial star formation in the cosmological code ART. We find that the effects of stellar feedback on the amount of baryons and metals within the first galaxies depend strongly on galaxy mass. For the lowest-mass galaxies ($M_h$ \textasciitilde$10^6 \, \Msun$) our results are similar to those of \citet{bromm_etal03, whalen_etal08, wise_abel08, wise_etal12}, with gas and metals often being driven well beyond the virial radius of the Pop III star's host galaxy. For more massive galaxies ($M_h \geq 10^7 \, \Msun $), however, a single PISN is not effective in evacuating the galactic ISM, as suggested by \citet{wise_cen09}. Feedback from Pop III stars does not typically inject enough energy into the massive halos to permanently photo-evaporate the gas, and drive metal-rich outflows past the virial radius. While Pop III stars can temporarily expel gas and quench star formation, the ISM begins to replenish soon after the SN explosion, as accretion from filaments at this epoch is very fast. All galaxies considered in our analysis with at least $M_h \approx 3 \times 10^6 \, \Msun$, and some which are even less massive, appear to have more than 50\% of the universal baryon fraction restored 100 Myr after the first Pop III supernova event. Metals are ejected anisotropically, and can travel relatively longer distances through the diffuse IGM in directions perpendicular to the dense filaments which feed galactic accretion. This means that it typically takes more time for the ejected metals to be re-accreted into the galaxy, but we have demonstrated that this re-accretion does frequently occur, even in low-mass galaxies. 

The aforementioned dividing line of $M_h \approx 3 \times 10^6 \, \Msun$ is important for determining whether the energy injection from the supernova at the end of the star's life can expel gas and metals out to a significant distance. The concept of a dividing line between early galaxies that suffer from significant blowout from those that do not has been considered in prior work (e.g. \citealt{ciardi_etal00, ricotti_etal02b}). However, our results point to a considerably lower threshold than what had been expected, as all but the very first galaxies are apparently robust to PISN feedback when continued accretion from filaments and the fallback of ejecta into the growing galaxy is considered. The strength of this conclusion is bolstered by our use of a very strong feedback model for Pop III stars (even in our fiducial runs). In addition, the first stars may have a lower characteristic mass \citep{greif_etal11}, which would make PISNe less frequent and the feedback effects would be further marginalized \citep{ritter_etal12}. 

In order for simulations to capture the full range of relevant effects from Pop III star formation, resolving halos around $10^6 \, \Msun$ with a sufficiently large number of particles is critical. With insufficient resolution (less than 1000 DM particles for $10^6 \, \Msun$ halos), all galaxies seem to reach $M > 10^7 \, \Msun$ without having yet formed a star. Since these galaxies are already beyond the $M_h \approx 3 \times 10^6 \, \Msun$ dividing line, they display few disruptive effects from Pop III feedback. Aggressive super-Lagrangian refinement may help resolve star formation in halos of lower mass, but requires a prohibitively large number of computations. A more practical approach is to begin simulations with sufficiently high resolution in the initial conditions.

\acknowledgements

A.L.M. acknowledges the support of the Rackham pre-Doctoral Fellowship awarded by The University of Michigan. O.Y.G. was supported in part by NSF grant AST-0708087 and NASA grant NNX12AG44G. This work was supported in part by the DOE at Fermilab and by the NSF grant AST-0708154. M.Z. is in part supported by a 985 grant from Peking University. We thank the anonymous referee for a thorough and insightful report. We also thank Alexander Heger for discussion of the evolution of a 100 $\Msun$ star, and John Wise, Andrey Kravtsov, Milos Milosavljevic, Michael Anderson, Wen-Hsin Hsu, and Doug Rudd for various constructive suggestions. 

\makeatletter\@chicagotrue\makeatother

\bibliography{FS}

\begin{thebibliography}{27}
\expandafter\ifx\csname natexlab\endcsname\relax\def\natexlab#1{#1}\fi

\bibitem[Abel et al.(2000)]{abel_etal00} Abel, T., Bryan, G.~L., 
\& Norman, M.~L.\ 2000, \apj, 540, 39 

\bibitem[Abel et al.(2002)]{abel_etal02} Abel, T., Bryan, G.~L., 
\& Norman, M.~L.\ 2002, Science, 295, 93 

\bibitem[Abel et al.(2007)]{abel_etal07} Abel, T., Wise, J.~H., 
\& Bryan, G.~L.\ 2007, \apjl, 659, L87 

\bibitem[Ahn et al.(2012)]{ahn_etal12} Ahn, K., Iliev, I.~T., 
Shapiro, P.~R., et al.\ 2012, \apjl, 756, L16 

\bibitem[Alvarez et al.(2006)]{alvarez_etal06} Alvarez, M.~A., Bromm, 
V., \& Shapiro, P.~R.\ 2006, \apj, 639, 621 

\bibitem[Alvarez et al.(2012)]{alvarez_etal12} Alvarez, M.~A., 
Finlator, K., \& Trenti, M.\ 2012, \apjl, 759, L38 

\bibitem[Bromm et al.(1999)]{bromm_etal99} Bromm, V., Coppi, P.~S., 
\& Larson, R.~B.\ 1999, \apjl, 527, L5 

\bibitem[Bromm et al.(2001)]{bromm_etal01} Bromm, V., Ferrara, A., 
Coppi, P.~S., \& Larson, R.~B.\ 2001, \mnras, 328, 969 

\bibitem[Bromm et al.(2002)]{bromm_etal02} Bromm, V., Coppi, P.~S., \& Larson, R.~B.\ 2002, \apj, 564, 23 

\bibitem[Bromm et al.(2003)]{bromm_etal03} Bromm, V., Yoshida, N., \& Hernquist, L.\ 2003, \apjl, 596, L135 

\bibitem[Ceverino 
\& Klypin(2009)]{ceverino_klypin09} Ceverino, D., \& Klypin, A.\ 2009, \apj, 695, 292 

\bibitem[Ciardi et al.(2000)]{ciardi_etal00} Ciardi, B., Ferrara, A., 
\& Abel, T.\ 2000, \apj, 533, 594 

\bibitem[Clark et al.(2011)]{clark_etal11} Clark, P.~C., Glover, 
S.~C.~O., Klessen, R.~S., \& Bromm, V.\ 2011, \apj, 727, 110 

\bibitem[Couchman 
\& Rees(1986)]{couchman_rees86} Couchman, H.~M.~P., \& Rees, M.~J.\ 1986, \mnras, 221, 53 


\bibitem[Gnedin 
\& Abel(2001)]{gnedin_abel01} Gnedin, N.~Y., \& Abel, T.\ 2001, New Astronomy, 6, 437 

\bibitem[Gnedin et al.(2008)]{gnedin_etal08} Gnedin, N.~Y., Kravtsov, 
A.~V., \& Chen, H.-W.\ 2008, \apj, 672, 765 

\bibitem[Gnedin et al.(2009)]{gnedin_etal09} Gnedin, N.~Y., Tassis, 
K., \& Kravtsov, A.~V.\ 2009, \apj, 697, 55 

\bibitem[Gnedin 
\& Kravtsov(2010)]{gnedin_kravtsov10} Gnedin, N.~Y., \& Kravtsov, A.~V.\ 2010, \apj, 714, 287 

\bibitem[Gnedin 
\& Kravtsov(2011)]{gnedin_kravtsov11} Gnedin, N.~Y., \& Kravtsov, A.~V.\ 2011, \apj, 728, 88 

\bibitem[Gnedin et al.(2011)]{gnedin_etal11} Gnedin, N.~Y., Kravtsov, 
A.~V., \& Rudd, D.~H.\ 2011, \apjs, 194, 46 

\bibitem[Greif et al.(2010)]{greif_etal10} Greif, T.~H., Glover, 
S.~C.~O., Bromm, V., \& Klessen, R.~S.\ 2010, \apj, 716, 510 

\bibitem[Greif et al.(2011)]{greif_etal11} Greif, T.~H., Springel, 
V., White, S.~D.~M., et al.\ 2011, \apj, 737, 75 

\bibitem[Greif et al.(2012)]{greif_etal12} Greif, T.~H., Bromm, V., 
Clark, P.~C., et al.\ 2012, \mnras, 424, 399 

\bibitem[Heger 
\& Woosley(2002)]{heger_woosley02} Heger, A., \& Woosley, S.~E.\ 2002, \apj, 567, 532 

\bibitem[Heger 
\& Woosley(2010)]{heger_woosley10} Heger, A., \& Woosley, S.~E.\ 2010, \apj, 724, 341 

\bibitem[Hopkins et al.(2011)]{hopkins_etal11} Hopkins, P.~F., 
Quataert, E., \& Murray, N.\ 2011, \mnras, 417, 950 

\bibitem[Hosokawa et al.(2011)]{hosokawa_etal11} Hosokawa, T., Omukai, K., Yoshida, N., \& Yorke, H.~W.\ 2011, Science, 334, 1250 

\bibitem[Hummels \& Bryan(2012)]{hummels_bryan12} Hummels, C.~B., \& Bryan, G.~L.\ 2012, \apj, 749, 140 

\bibitem[Johnson et al.(2007)]{johnson_etal07} Johnson, J.~L., Greif, 
T.~H., \& Bromm, V.\ 2007, \apj, 665, 85 

\bibitem[Kravtsov et al.(1997)]{kravtsov_etal97} Kravtsov, A.~V., 
Klypin, A.~A., \& Khokhlov, A.~M.\ 1997, \apjs, 111, 73 

\bibitem[Kravtsov(1999)]{kravtsov99} Kravtsov, A.~V.\ 1999, 
Ph.D.~Thesis, New Mexico State University

\bibitem[Kravtsov(2003)]{kravtsov03} Kravtsov, A.~V.\ 2003, \apjl, 
590, L1 

\bibitem[Krumholz et al.(2012)]{krumholz_etal12} Krumholz, M.~R., 
Dekel, A., \& McKee, C.~F.\ 2012, \apj, 745, 69 

\bibitem[Leitherer et al.(1999)]{leitherer_etal99} Leitherer, C., 
Schaerer, D., Goldader, J.~D., et al.\ 1999, \apjs, 123, 3 

\bibitem[Machacek et al.(2001)]{machacek_etal01} Machacek, M.~E., 
Bryan, G.~L., \& Abel, T.\ 2001, \apj, 548, 509 

\bibitem[Maio et al.(2011)]{maio_etal11} Maio, U., Khochfar, S., 
Johnson, J.~L., \& Ciardi, B.\ 2011, \mnras, 414, 1145 

\bibitem[Mesinger et al.(2009)]{mesinger_etal09} Mesinger, A., Bryan, 
G.~L., \& Haiman, Z.\ 2009, \mnras, 399, 1650 

\bibitem[Miller 
\& Scalo(1979)]{miller_scalo79} Miller, G.~E., \& Scalo, J.~M.\ 1979, \apjs, 41, 513 

\bibitem[Muratov et al.(2013a)]{muratov_etal13} Muratov, A.~L., Gnedin, 
O.~Y., Gnedin, N.~Y., \& Zemp, M.\ 2013, \apj, 773, 19 

\bibitem[O'Shea 
\& Norman(2007)]{oshea_norman07} O'Shea, B.~W., \& Norman, M.~L.\ 2007, \apj, 654, 66 

\bibitem[O'Shea 
\& Norman(2008)]{oshea_norman08} O'Shea, B.~W., \& Norman, M.~L.\ 2008, \apj, 673, 14 

\bibitem[Omukai et al.(2005)]{omukai_etal05} Omukai, K., Tsuribe, T., 
Schneider, R., \& Ferrara, A.\ 2005, \apj, 626, 627 

\bibitem[Ricotti et al.(2002a)]{ricotti_etal02a} Ricotti, M., Gnedin, 
N.~Y., \& Shull, J.~M.\ 2002, \apj, 575, 33 

\bibitem[Ricotti et al.(2002b)]{ricotti_etal02b} Ricotti, M., Gnedin, 
N.~Y., \& Shull, J.~M.\ 2002, \apj, 575, 49 

\bibitem[Ricotti 
\& Ostriker(2004)]{ricotti_ostriker04} Ricotti, M., \& Ostriker, J.~P.\ 2004, \mnras, 350, 539 

\bibitem[Ricotti et al.(2008)]{ricotti_etal08} Ricotti, M., Gnedin, 
N.~Y., \& Shull, J.~M.\ 2008, \apj, 685, 21 

\bibitem[Ritter et al.(2012)]{ritter_etal12} Ritter, J.~S., 
Safranek-Shrader, C., Gnat, O., Milosavljevi{\'c}, M., 
\& Bromm, V.\ 2012, \apj, 761, 56 

\bibitem[Rudd et al.(2008)]{rudd_etal08} Rudd, D.~H., Zentner, 
A.~R., \& Kravtsov, A.~V.\ 2008, \apj, 672, 19 

\bibitem[Safranek-Shrader et al.(2012)]{safranek_etal12} 
Safranek-Shrader, C., Agarwal, M., Federrath, C., et al.\ 2012, \mnras, 
426, 1159 

\bibitem[Schaerer(2002)]{schaerer02} Schaerer, D.\ 2002, \aap, 382, 28 

\bibitem[Sirko(2005)]{sirko05} Sirko, E.\ 2005, \apj, 634, 728 

\bibitem[Smith et al.(2009)]{smith_etal09} Smith, B.~D., Turk, 
M.~J., Sigurdsson, S., O'Shea, B.~W., 
\& Norman, M.~L.\ 2009, \apj, 691, 441 

\bibitem[Stacy et al.(2010)]{stacy_etal10} Stacy, A., Greif, T.~H., 
\& Bromm, V.\ 2010, \mnras, 403, 45 

\bibitem[Stacy et al.(2012)]{stacy_etal12} Stacy, A., Greif, T.~H., 
\& Bromm, V.\ 2012a, \mnras, 422, 290 

\bibitem[Stacy et al.(2013)]{stacy_etal13} Stacy, A., Greif, T.~H., 
Klessen, R.~S., Bromm, V., \& Loeb, A.\ 2013, \mnras, 431, 1470 



\bibitem[Tassis et al.(2008)]{tassis_etal08}Tassis, K., Kravtsov, 
A.~V., \& Gnedin, N.~Y.\ 2008, \apj, 672, 888 

\bibitem[Tegmark et al.(1997)]{tegmark_etal97} Tegmark, M., Silk, J., 
Rees, M.~J., et al.\ 1997, \apj, 474, 1 

\bibitem[Tumlinson 
\& Shull(2000)]{tumlinson_shull00} Tumlinson, J., \& Shull, J.~M.\ 2000, \apjl, 528, L65 

\bibitem[Turk et al.(2009)]{turk_etal09} Turk, M.~J., Abel, T., 
\& O'Shea, B.\ 2009, Science, 325, 601 

\bibitem[Whalen et al.(2004)]{whalen_etal04} Whalen, D., Abel, T., 
\& Norman, M.~L.\ 2004, \apj, 610, 14 

\bibitem[Whalen et al.(2008)]{whalen_etal08} Whalen, D., van Veelen, 
B., O'Shea, B.~W., \& Norman, M.~L.\ 2008, \apj, 682, 49 

\bibitem[Wise 
\& Abel(2008)]{wise_abel08} Wise, J.~H., \& Abel, T.\ 2008, \apj, 685, 40 

\bibitem[Wise 
\& Cen(2009)]{wise_cen09} Wise, J.~H., \& Cen, R.\ 2009, \apj, 693, 984 

\bibitem[Wise et al.(2012)]{wise_etal12} Wise, J.~H., Turk, M.~J., 
Norman, M.~L., \& Abel, T.\ 2012, \apj, 745, 50 

\bibitem[Woosley 
\& Heger(2012)]{woosley_heger12} Woosley, S.~E., \& Heger, A.\ 2012, \apj, 752, 32 

\bibitem[Yoshida et al.(2003)]{yoshida_etal03} Yoshida, N., Abel, T., 
Hernquist, L., \& Sugiyama, N.\ 2003, \apj, 592, 645 

\bibitem[Yoshida et al.(2004)]{yoshida_etal04} Yoshida, N., Bromm, V., \& Hernquist, L.\ 2004, \apj, 605, 579 

\bibitem[Yoshida et al.(2006)]{yoshida_etal06} Yoshida, N., Omukai, 
K., Hernquist, L., \& Abel, T.\ 2006, \apj, 652, 6 

\bibitem[Yoshida et al.(2007)]{yoshida_etal07} Yoshida, N., Omukai, 
K., \& Hernquist, L.\ 2007, \apjl, 667, L117 


\bibitem[Zemp et al.(2012)]{zemp_etal12} Zemp, M., Gnedin, O.~Y.,  Gnedin, N.~Y., \& Kravtsov, A.~V.\ 2012, \apj, 748, 54 


\end{thebibliography}

\end{document}